\documentclass[journal,12pt,onecolumn,letterpaper]{IEEEtran}
\usepackage{arxiv}
\usepackage{geometry}
\usepackage{times}
\usepackage{cite}
\usepackage{url}
\usepackage{graphicx}
\graphicspath{{Images_SimplyMime/}}
\usepackage{lscape}
\usepackage{subfigure}
\usepackage{rotating}
\usepackage{rotfloat}
\usepackage{xcolor}
\usepackage{amsmath}
\usepackage{amssymb}
\usepackage[linesnumbered,ruled,vlined]{algorithm2e}
\usepackage{pseudocode}
\usepackage{array}
\usepackage[english]{babel}
\usepackage{gensymb}
\usepackage{textcomp}
\usepackage{placeins}
\usepackage{balance}
\usepackage{booktabs}

\title{Blind Eye: Motion and Obstacle Detection Leveraging Wi-Fi}

\author{%
  
    Aditya Mitra \\
    Centre of Excellence, Artificial Intelligence \& Robotics (AIR),\\
    School of Computer Science and Engineering\\
    VIT-AP University, India \\
    \texttt{adityaarghya0@gmail.com}
\And

Anisha Ghosh\\
    Centre of Excellence, Artificial Intelligence \& Robotics (AIR),\\
    School of Computer Science and Engineering\\
    VIT-AP University, India \\
    \texttt{ghoshanisha2002@gmail.con}
\And
  Sibi Chakkaravarthy Sethuraman\\
    Centre of Excellence, Artificial Intelligence \& Robotics (AIR),\\
    School of Computer Science and Engineering\\
    VIT-AP University, India \\
    \texttt{sb.sibi@gmail.com} \\
\And
    Devi Priya V S \\
    Department of Computer Science and Engineering [Cyber Security]\\
    Dayanandha Sagar University, India \\	
   \texttt{vsdevipriya@gmail.com}
}

\begin{document}

    


\maketitle

\begin{abstract}
 Wireless Fidelity or Wi-Fi, has completely transfigured wireless networking by offering a smooth connection to the internet and networks, particularly when dealing with enclosed environments. As with the majority of wireless technology, it functions through radio communication. This makes it possible for Wi-Fi to operate effectively close to an Access Point. However, a device's ability to receive Wi-Fi signals can vary greatly. These discrepancies arise because of impediments or motions between the device and the access point. We have creatively used these variances as unique opportunities for applications that can be used to detect movement in confined areas. As this approach makes use of the current wireless infrastructure, no additional hardware is required. These applications could potentially be leveraged to enable sophisticated robots or enhance security systems.

\end{abstract}




\maketitle

\section{Introduction}

Conventional motion and obstacle detection systems detect movement or impediments in a particular environment using cameras, ultrasonic waves, and infrared sensors \cite{wang2021research,guo2023recent,kuribayashi2022corridor}. As they can identify variations in heat signatures, infrared sensors are useful for automated lighting and security systems. Ultrasonic sensors, frequently seen in cars for parking assistance, produce sound waves and measure the reflection to calculate the distance to objects. In applications like robotics and surveillance, cameras— frequently paired with image processing algorithms—provide visual data for motion and obstacle detection with high accuracy. These methods have been used for decades, although they often require specialized hardware and careful setup which can limit flexibility and elevate prices. Furthermore, the calibration and installation procedures can be intricate and require precise configuration to guarantee peak performance. 

Wi-Fi could drastically cut expenses in motion and obstacle detection systems by utilizing the prevailing wireless infrastructure. Wi-Fi-based systems employ software algorithms to analyze signal fluctuations brought on by movement, in contrast to traditional approaches that require specialized sensors or cameras. This method further reduces expenses by reducing hardware purchases, calibration, and maintenance requirements. Wi-Fi is also affordable for various settings because it can cover wide regions with a single access point, guaranteeing excellent scalability. IEEE 802.11x wireless communication technologies, referred to as Wi-Fi, employ devices like an Access Point and a Network Interface Card (NIC) with Wi-Fi capability and operate based on radio communication \cite{liu2020survey,wu2021hybrid}. Wi-Fi enabled NICs are embedded into the majority of smartphones, laptops, wearables, smart home appliances, etc. Most residential homes, workplaces, parks, and other locations deploy Wi-Fi to facilitate quick access to the internet \cite{javed2022future}. The built-in Wi-Fi networks in smartphones and other devices are sufficient for various types of networks. Among the several kinds of Wi-Fi networks are:
\begin{itemize}
    \item  Ad-Hoc: This method does not make use of a centralized access point. It is used for short-term access and is frequently made possible by setting up a hotspot on a laptop or smartphone.
    \item  Mesh Network: A decentralized network that provides coverage over a wide geographic region, frequently accomplished by using several Access Points or Repeaters that serve as nodes and communicate with one another.
    \item WiFi Direct: There are just two devices involved in this peer-to-peer network. It is frequently used for tasks like file sharing, printing, wireless monitors, and more.
\end{itemize}
Wi-Fi networks vary greatly depending on the operating frequency and speed as defined by the IEEE 802.11x specifications. The frequency bands that Wi-Fi networks often use are 2.4 GHz, 5 GHz, and more recently, 6 GHz.

\subsection{Contribution in this research}

We aim to develop and implement cost-effective and, scalable methods for physical reconnaissance and stealth purposes. We contribute the following in that endeavor:

\begin{itemize}
\item We introduce a pioneering innovative approach for detecting movement in confined areas by examining Wi-Fi signal strength (RSSI) variations caused by motion and physical obstacles.
\item  This approach harnesses the prevailing Wi-Fi infrastructure to detect passive motion without the need for specific sensing hardware, making it a scalable and economical solution.

\end{itemize}
\section{State-of-the-Art}

Wi-Fi-based obstacle localization in a specific area has been the subject of several earlier studies. Deep learning techniques were applied to the Wi-Fi signals transmitted and received by three customized six DBi copper dipole antennas each in a study \cite{wang2019person} by Wang, F. et al. To train a deep learning algorithm that accurately detects the humans in the path, the researchers annotated the Wi-Fi signals against image recognition performed by cameras. Super-resolution techniques to accurately locate a person utilizing Wi-Fi in an indoor environment were proposed in the work \cite{kotaru2015spotfi}. However, it is ineffective for passive reconnaissance and necessitates a Wi-Fi enabled equipment. An array of Wi-Fi access points or transmitters to view the variations in Wi-Fi signals across a concrete wall is used to create a multi-antenna device that can see through walls \cite{adib2013see}. Yet, it was incompatible with passive reconnaissance and required the target to have a Wi-Fi-enabled gadget on hand.

There were studies available for locating a person in 2D or 3D space inside a Wi-Fi range. In some studies, camera feeds are utilized to annotate Wi-Fi data, which is then used to display people in three dimensions. RSSI-based indoor localization of Wi-Fi devices in a specific region is the subject of several studies \cite{zhu2013rssi,yen20223,wang2017research,dubey2021enhanced,singh2021machine,sadowski2018rssi}, which entails locating a device without detecting unexpected motion. Some of them employ machine learning methods, which makes them unsuitable for usage with portable smart home appliances.  According to research by Lui, G. et al. \cite{lui2011differences}, RSSI fingerprinting may be unreliable at low bandwidths, highlighting its limits.
So, we suggest Blind-Eye, which focuses on motion and obstruction detection and does not require the subject to carry a Wi-Fi device.  It focuses on motion detection rather than detecting devices within the Wi-Fi range. Table \ref{Comparitive Study of State-of-the-art with Blind Eye} summarizes the comparative analysis of the state-of-the-art with Blind Eye.

\begin{table*}[]
\centering
\caption{Comparative Analysis of Blind Eye and State-of-the-Art}
\label{Comparitive Study of State-of-the-art with Blind Eye}
\renewcommand{\arraystretch}{1} 
\resizebox{\textwidth}{!}{%
\large 
\begin{tabular}{|l|l|l|l|l|}
\hline
\multicolumn{1}{|c|}{\textbf{Features}}                                         & \textbf{Person-in-Wifi \cite{wang2019person}  }                                                                                                                                               & \textbf{Spotfi \cite{kotaru2015spotfi}   }                                                                     & \begin{tabular}[c]{@{}l@{}} \textbf{See through walls}\\\textbf{with WiFi \cite{adib2013see} } \end{tabular}                                                                                     &\textbf{ Blind Eye (The proposed work) }                                                                                                    \\ \hline
\begin{tabular}[c]{@{}l@{}}Specialized \\ Hardware\end{tabular}                 & \begin{tabular}[c]{@{}l@{}}Not required, but it needs a \\ camera for training and a \\ machine capable of running \\ Deep Learning models\end{tabular} & Not Required                                                                                               & \begin{tabular}[c]{@{}l@{}}Requires a specialized \\ Wi-Fi device\end{tabular}                                      & Not Required                                                                                                  \\ \hline
\begin{tabular}[c]{@{}l@{}}Passive \\ Reconnaissance \\ Capability\end{tabular} & No                                                                                                                                                      & No                                                                                                         & No                                                                                                                  & Yes                                                                                                           \\ \hline
\begin{tabular}[c]{@{}l@{}}Environmental \\ Dependence\end{tabular}             & \begin{tabular}[c]{@{}l@{}}Requires a clear camera \\ view and controlled \\ lighting during training\end{tabular}                                      & \begin{tabular}[c]{@{}l@{}}Performs well \\ in most \\ environments\end{tabular}                           & \begin{tabular}[c]{@{}l@{}}Limited to environments \\ conducive to Wi-Fi \\ signal reflection analysis\end{tabular} & \begin{tabular}[c]{@{}l@{}}Minimal environmental \\ restrictions, works in \\ diverse conditions\end{tabular} \\ \hline
\begin{tabular}[c]{@{}l@{}}Targets Carry \\ Wi-Fi \\ Devices\end{tabular}       & Yes                                                                                                                                                     & Yes                                                                                                        & Yes                                                                                                                 & \begin{tabular}[c]{@{}l@{}}Detect and track targets \\ without them carrying\\ Wi-Fi devices\end{tabular}     \\ \hline
Accuracy                                                                        & \begin{tabular}[c]{@{}l@{}}High, dependent on camera\\ quality and model training\end{tabular}                                                          & \begin{tabular}[c]{@{}l@{}}Moderate to High, \\ depends on signal \\ processing accuracy\end{tabular}      & \begin{tabular}[c]{@{}l@{}}High, where Wi-Fi \\ signals are optimized.\end{tabular}                                 & \begin{tabular}[c]{@{}l@{}}High, especially for \\ stationary/slow-moving \\ targets\end{tabular}             \\ \hline
\begin{tabular}[c]{@{}l@{}}Range of \\ Operation\end{tabular}                   & \begin{tabular}[c]{@{}l@{}}Limited to the camera's \\ field of view during \\ training\end{tabular}                                                     & \begin{tabular}[c]{@{}l@{}}Limited to Wi-Fi \\ signal coverage\end{tabular}                                & \begin{tabular}[c]{@{}l@{}}Limited to Wi-Fi \\ signal coverage\end{tabular}                                         & \begin{tabular}[c]{@{}l@{}}Limited to the Wi-Fi coverage, depends on signal \\ processing setup and\\ conditions\end{tabular}         \\ \hline
Scalability                                                                     & \begin{tabular}[c]{@{}l@{}}Low, as adding more \\ targets requires additional \\ training\end{tabular}                                                  & \begin{tabular}[c]{@{}l@{}}Moderate, scales \\ with Wi-Fi network \\ size\end{tabular}                     & \begin{tabular}[c]{@{}l@{}}Low, dependent on \\ specialized devices\end{tabular}                                    & \begin{tabular}[c]{@{}l@{}}High, works well across \\ larger areas with minimal\\ tuning\end{tabular}         \\ \hline
\begin{tabular}[c]{@{}l@{}}Real-Time \\ Capability\end{tabular}                 & \begin{tabular}[c]{@{}l@{}}Limited, as training phases\\ take time and updates\\ aren't instantaneous.\end{tabular}                                     & \begin{tabular}[c]{@{}l@{}}Moderate, operates\\ in near real-time\end{tabular}                             & \begin{tabular}[c]{@{}l@{}}High, optimized for \\ real-time detection\end{tabular}                                  & \begin{tabular}[c]{@{}l@{}}High, capable of real-time \\ analysis without delays\end{tabular}                 \\ \hline
\begin{tabular}[c]{@{}l@{}}Cost of \\ Implementation\end{tabular}               & \begin{tabular}[c]{@{}l@{}}Moderate, primarily due to \\ machine learning resources\end{tabular}                                                        & \begin{tabular}[c]{@{}l@{}}Low, leverages \\ existing Wi-Fi \\ \textbackslash{}infrastructure\end{tabular} & \begin{tabular}[c]{@{}l@{}}High, owing to \\ specialized \\ hardware costs\end{tabular}                             & \begin{tabular}[c]{@{}l@{}}Low, no additional \\ equipment required\end{tabular}                              \\ \hline
\end{tabular}
}
\end{table*}

 \section{Proposed Methodology}
Typically, two Wi-Fi networks operating in adjacent geographic areas on the same frequency band use distinct, non-overlapping channels. This guarantees minimal interference from one network to another \cite{omar2016survey}. Wi-Fi receivers can individually identify each Wi-Fi network since they are on distinct, non-overlapping channels. Each AP's unique MAC address, known as the Basic Service Set Identifier or BSSI, is used to identify each network. The signal power of AP is typically measured in milliwatts (mW), and the RSSI (Received Signal Strength Indicator) is expressed in decibels (dBm) relative to 1 mW \cite{arigye2022rssi}.  

\begin{equation}
RSSI \: (dBm)=10 * log_{10}(Signal \: Power \: in\: mW) 
\end{equation}

\begin{figure}[H] 
    \centering
    \includegraphics[width=0.5\textwidth]{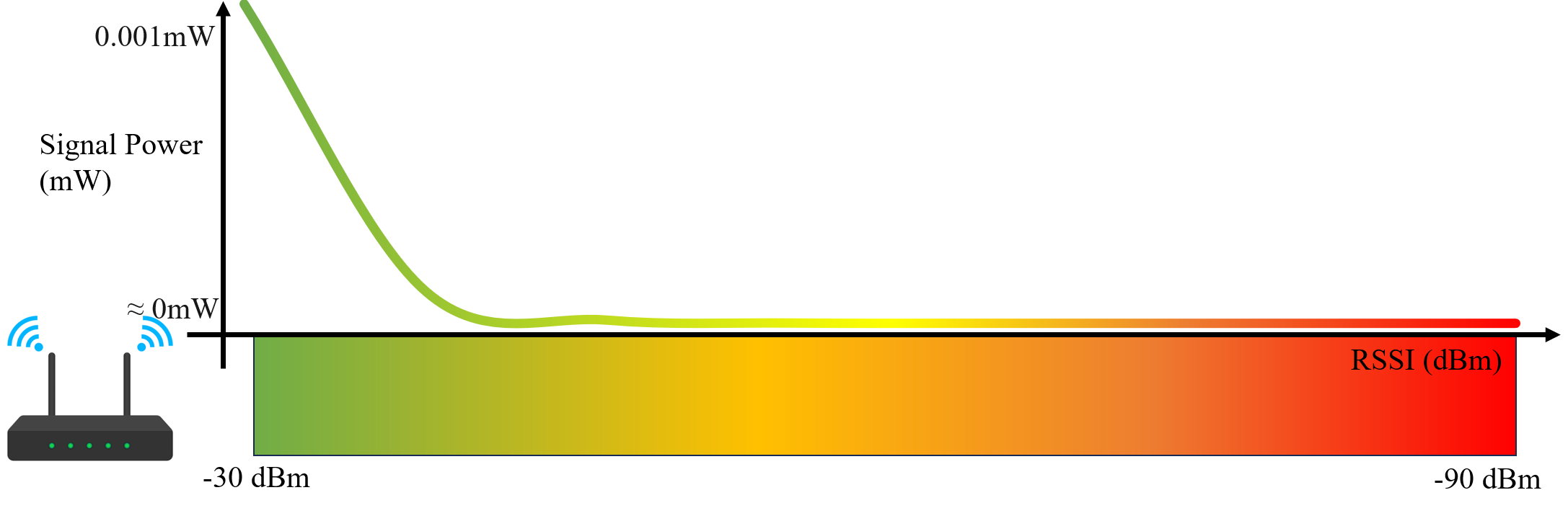} 
    \caption{Signal Power \textit{v/s} RSSI }
    \label{fig:Signal Power v/s RSSI}
\end{figure}

While a weak signal might be at -90 dBm, a strong signal might be around -30 dBm. The relationship between the RSSI values and the signal power is depicted in Figure \ref{fig:Signal Power v/s RSSI}. A strong signal or high RSSI value is indicated by the color green, whereas a weak signal or lower RSSI value is indicated by the color red. The remaining figures in this paper use the same color scheme.
The signal power drops when an impediment blocks the path between the AP and the receiver, changing the RSSI value. The possible causes for the signal power change can include \cite{bao2024addressing}:
\begin{itemize}
    \item  Attenuation: This is characterized by a progressive loss of strength and frequently occurs as the receiver's distance from the AP varies. The steady decline in signal intensity brought on by attenuation is depicted in Figure \ref{Signal Intensity Variation Caused by Attenuation}.
    \begin{figure}[htbp] 
    \centering
    \includegraphics[width=0.3\textwidth, height=4cm]{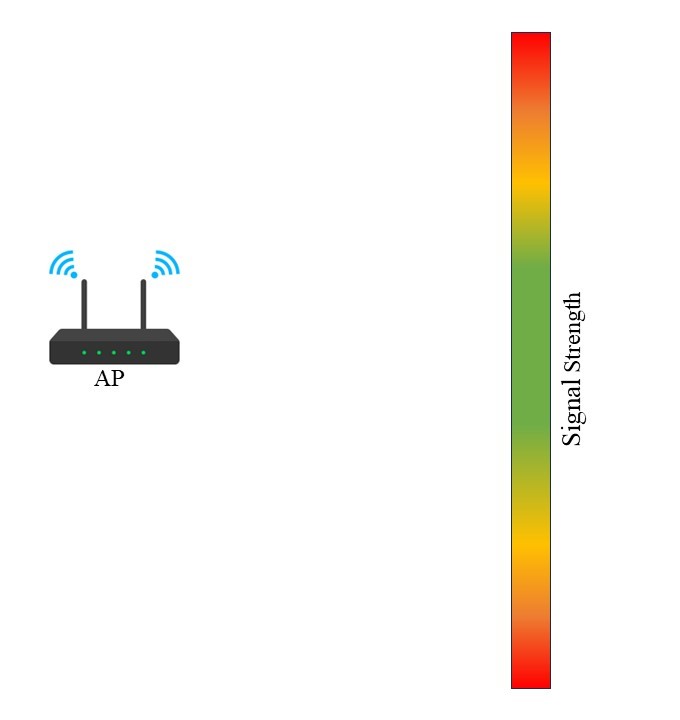} 
    \caption{Signal Intensity Variation Caused by Attenuation}
    \label{Signal Intensity Variation Caused by Attenuation}
\end{figure}
\item Shadowing: This is characterized by a noticeable decrease in signal intensity brought on by substantial physical barriers, such as walls or concrete, that hinder the AP and receiver's line of sight (LOS).
  \begin{figure}[H] 
    \centering
    \includegraphics[width=0.3\textwidth, height=4cm]{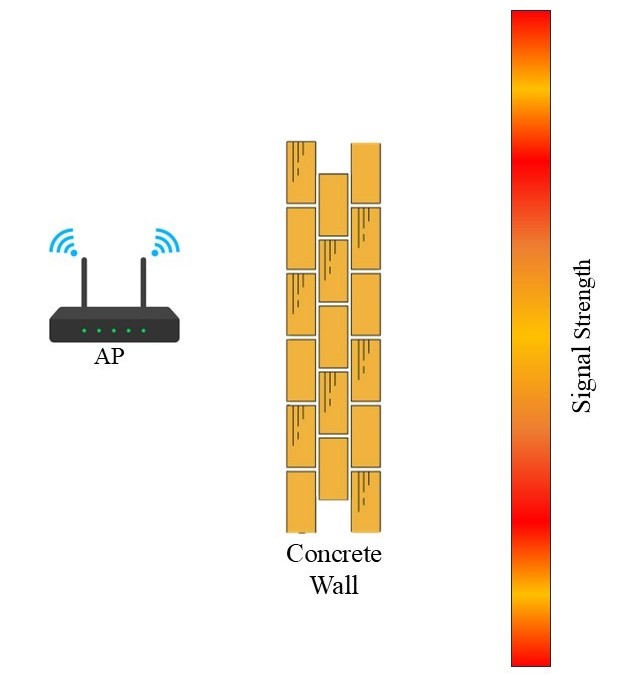} 
    \caption{Signal Intensity Variation Caused by Shadowing}
    \label{Signal Intensity Variation Caused by Shadowing}
\end{figure}
\item Absorption: This is apparent by a signal loss as the material it passes through is absorbed. Electromagnetic waves are absorbed by metal, water, and even human tissue.
  \begin{figure}[H] 
    \centering
    \includegraphics[width=0.3\textwidth, height=4cm]{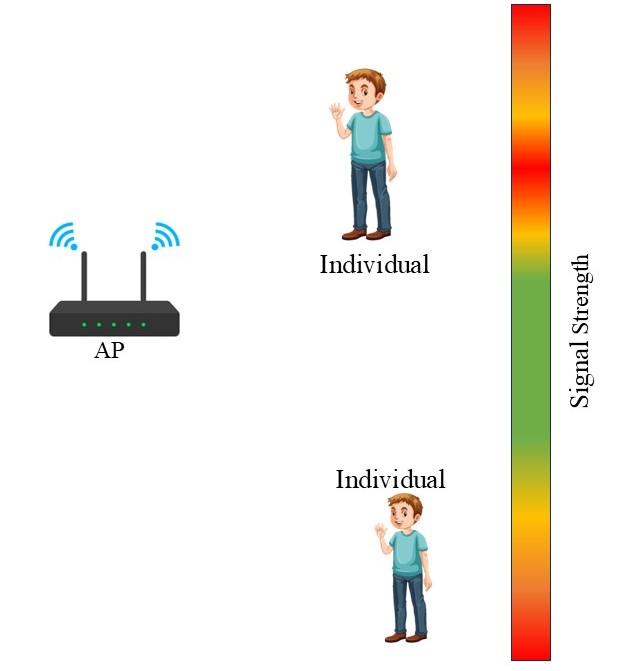} 
    \caption{Signal Intensity Variation Caused by Absorption}
    \label{Signal Intensity Variation Caused by Absorption}
\end{figure}
\end{itemize}
Diffraction, scattering, and multipath fading are some additional causes of signal power changes, but they are not as significant as attenuation, shadowing, or absorption. 

In view of the distinctive manner in which the RSSI values respond to motion and impediments, we have used this approach to detect motion without requiring a lot of expensive hardware. Using a Commercial-Off-The-Shelf (COTS) Wi-Fi access station, one can read the Wi-Fi RSSI values and detect motion using a standard smartphone, laptop, or smart home device. The proposed approach for motion detection entails configuring a Wi-Fi AP using a COTS router or an ad hoc device, such as a smartphone, laptop, or smart home device, and utilizing the receiver to track the signal strength. A motion or obstruction between the transmitter and the receiver may be indicated by a significant loss in signal strength without a change in the transmitter or receiver's distance or orientation. It has been noted that the approach provides better visibility than conventional motion sensors and operates accurately even through physical barriers like concrete walls.
\section{Experimental Setup and Results}
The experimental setup employed a laptop with an Intel Wi-Fi AX211 Wi-Fi card as the receiver and a COTS router, the TP-Link TL-WR845N with three 5 dBi high gain antennas as the transmitter. Ubuntu 22.04 LTS and a custom-written Python script were configured on the laptop to report and analyze the signal levels.
 \begin{figure}[H] 
    \centering
    \includegraphics[width=0.45\textwidth, height=5cm]{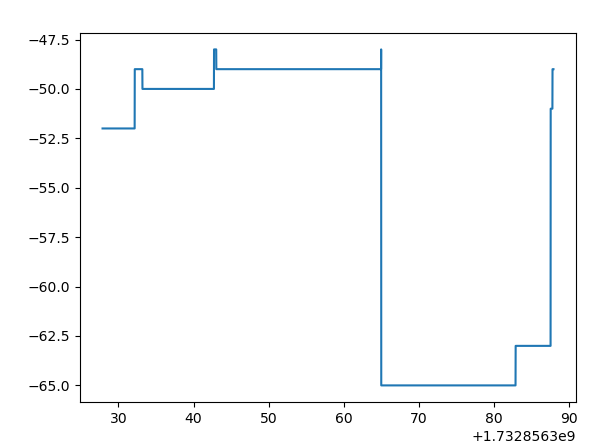} 
    \caption{Single Obstruction between Transmitter and Receiver}
    \label{Single obstruction between transmitter and receiver}
\end{figure}

When the AP and the receiver are positioned 2.5 meters apart in LOS, Figure \ref{Single obstruction between transmitter and receiver} illustrates how the RSSI values change when a human impediment falls between them. When no obstruction existed, the average RSSI value was discovered to be approximately -50 dBm. Signal strength, as measured by the RSSI value, has been found to decrease to -65 dBm when a barrier is placed between the AP and the receiver.

\begin{figure}[H] 
    \centering
    \includegraphics[width=0.4\textwidth, height=5cm]{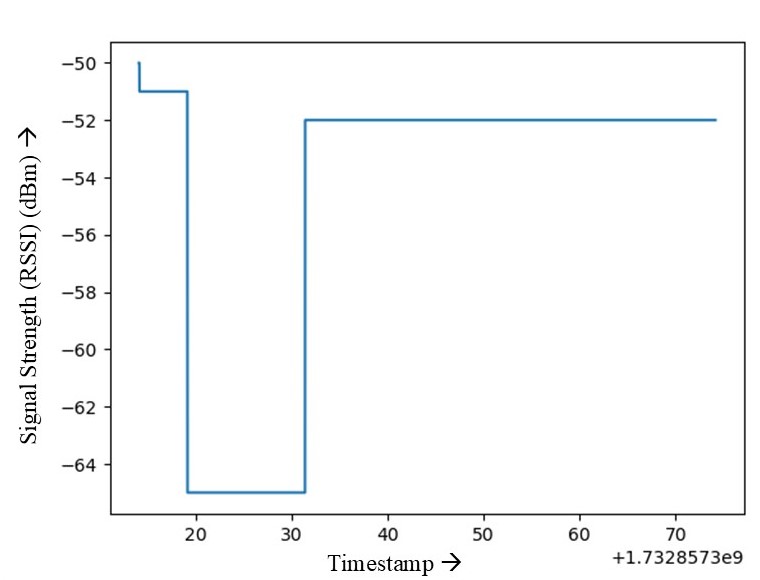} 
    \caption{Obstruction between Transmitter and Receiver at LOS=5.3m}
    \label{Obstruction between transmitter and receiver at LOS=5.3m}
\end{figure}

The AP and the receiver were maintained in LOS at a distance of 5.3 meters in another trial. When there was a clear LOS, the average RSSI value was approximately -52 dBm; however, when an obstacle, such as a human, crossed it, it decreased to approximately -64 dBm. The resulting plot is displayed in Figure \ref{Obstruction between transmitter and receiver at LOS=5.3m}.

\begin{figure}[H] 
    \centering
    \includegraphics[width=0.45\textwidth, height=4.5cm]{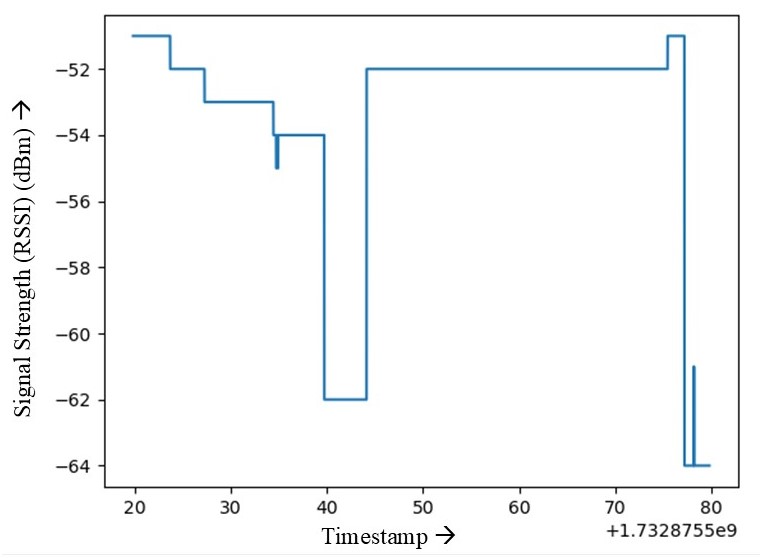} 
    \caption{Plot of Result when two Different Obstruction in LOS}
    \label{Two different obstructions}
\end{figure}
The average RSSI value was around -53 dBm when there was no obstruction, but dropped to -62 dBm and -64 dBm when different people were introduced in between.The AP and the receiver were placed approximately 5.3 meters apart in LOS. The plot of the same is displayed in Figure \ref{Two different obstructions}.
\begin{figure}[H] 
    \centering
    \includegraphics[width=0.45\textwidth, height=4.5cm]{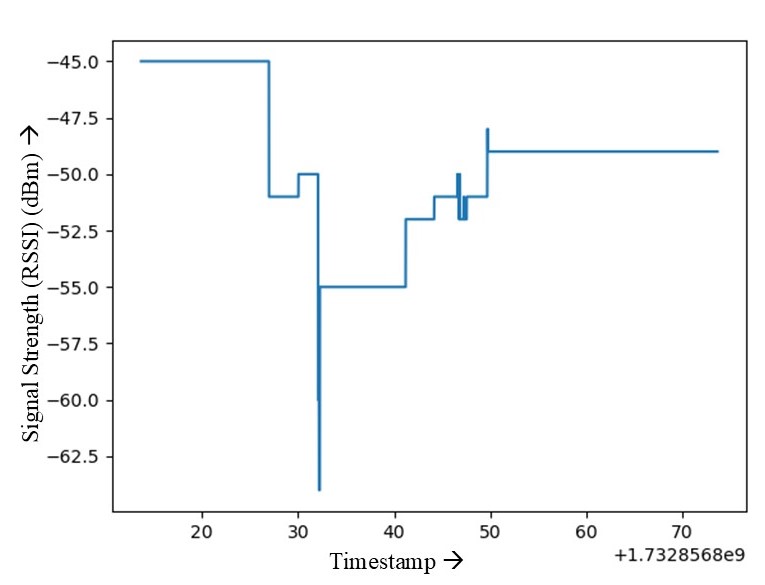} 
    \caption{Momentary drop in signal when obstruction moves fast}
    \label{Momentary drop in signal when obstruction moves fast}
\end{figure}

In a subsequent trial performed, the human traveled quickly and continuously over the LOS. The RSSI value in this instance fell to -61 dBm at that same moment, although the typical RSSI value was approximately -50 dBm. This made sure that even if the object moves quickly, the experimental setup can detect it. Figure \ref{Momentary drop in signal when obstruction moves fast} illustrates the brief decline in the signal values. 

\begin{figure}[H] 
    \centering
    \includegraphics[width=0.45\textwidth, height=4.5cm]{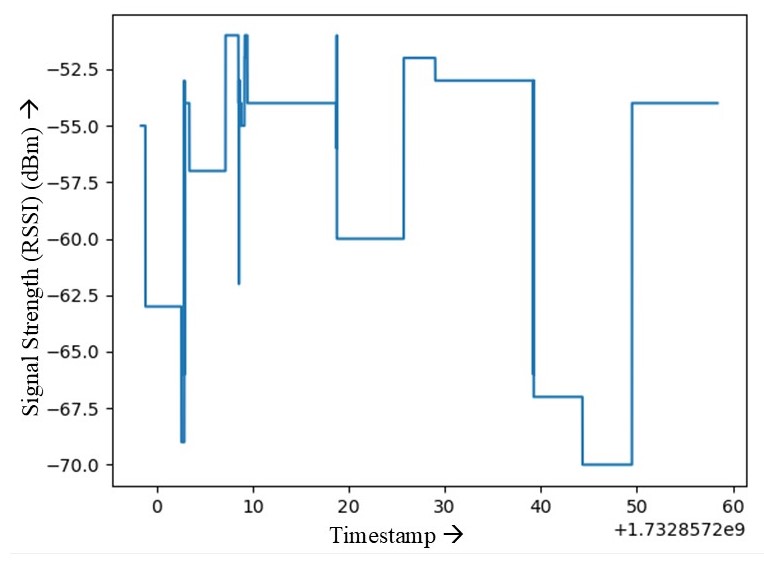} 
    \caption{Signal across a wall}
    \label{Signal across a wall}
\end{figure}

To discover if the setup could see through walls, more tests were carried out. The receiver and the AP were positioned 4m apart, on opposite sides of a concrete wall. A person is permitted to cross the path between the receiver and the AP. He walked twice on the side of the AP and twice on the side of the receiver. In the absence of the person, the average RSSI value was observed to be approximately -55 dBm. The RSSI value decreased to around -60 dBm when the person walked on the receiver's side and to about -70 dBm when the person walked on the AP's side. Figure \ref{Signal across a wall} illustrates the configuration and wall orientation.

\begin{figure}[H] 
    \centering
    \includegraphics[width=0.45\textwidth, height=4.5cm]{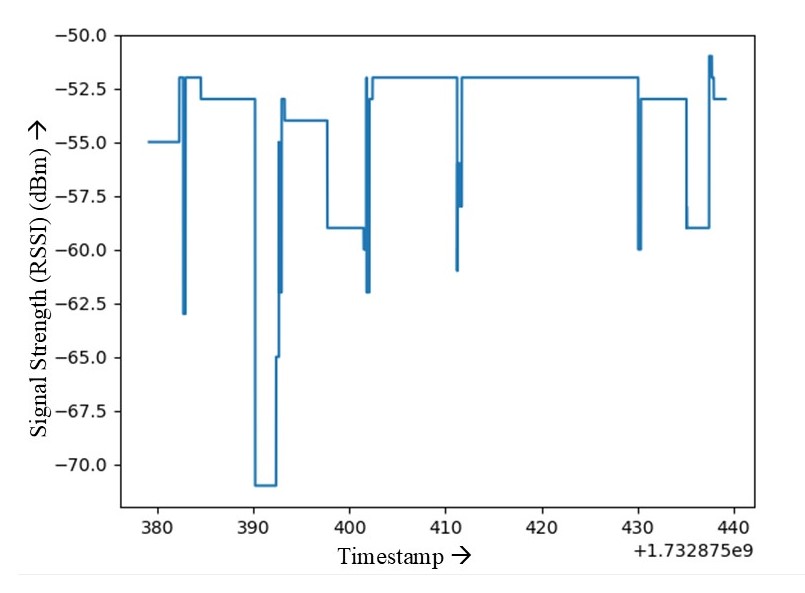} 
    \caption{Signal when Walking across the Wall}
    \label{Signal when walking across the wall}
\end{figure}
\begin{figure}[h] 
    \centering
    \includegraphics[width=0.4\textwidth]{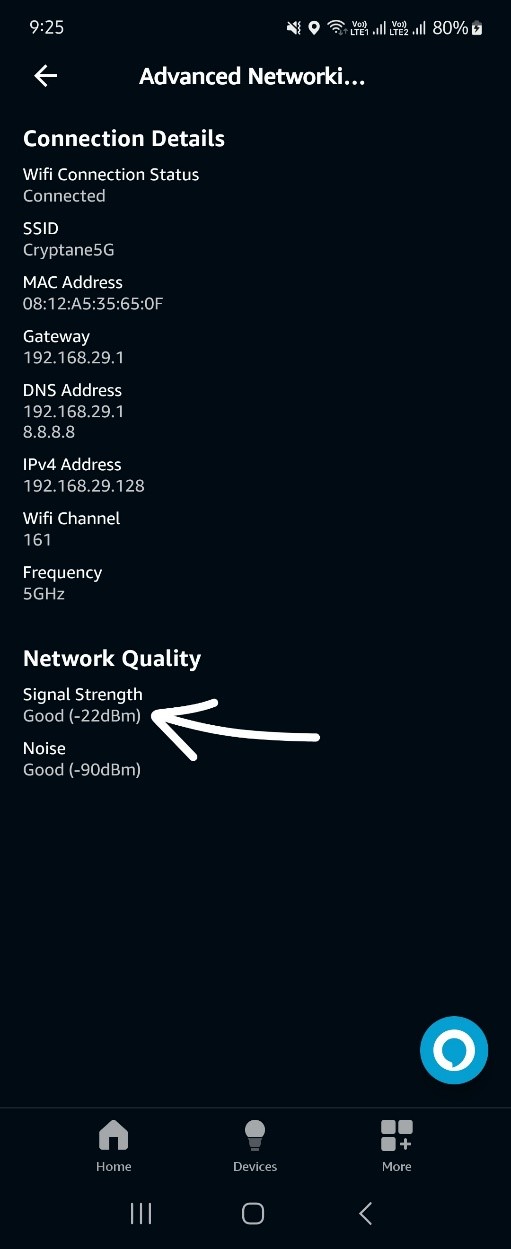} 
    \caption{RSSI Value Reported by Echo Dot}
    \label{RSSI value reported by Echo Dot}
\end{figure}

In a related test, the subject repeatedly crossed the receiver without pausing. The motion that resulted from the transient blockages was precisely detected by the receiver. The RSSI value was approximately -51 dBm on average. The RSSI value decreased to -61 dBm when a human crossed the setup on the receiver's side, then again to -70 dBm when the receiver crossed on the AP's side. The ensuing changes in the RSSI values are displayed in Figure \ref{Signal when walking across the wall}.
Furthermore, any smart home appliance, such as smart speakers, has been shown to produce comparable outcomes. The values supplied by an Amazon Echo Dot, a multipurpose smart speaker, are displayed in Figure \ref{RSSI value reported by Echo Dot}.

\section{Usecase Scenarios}
\begin{figure}[h] 
    \centering
    \includegraphics[width=0.4\textwidth, height=5cm]{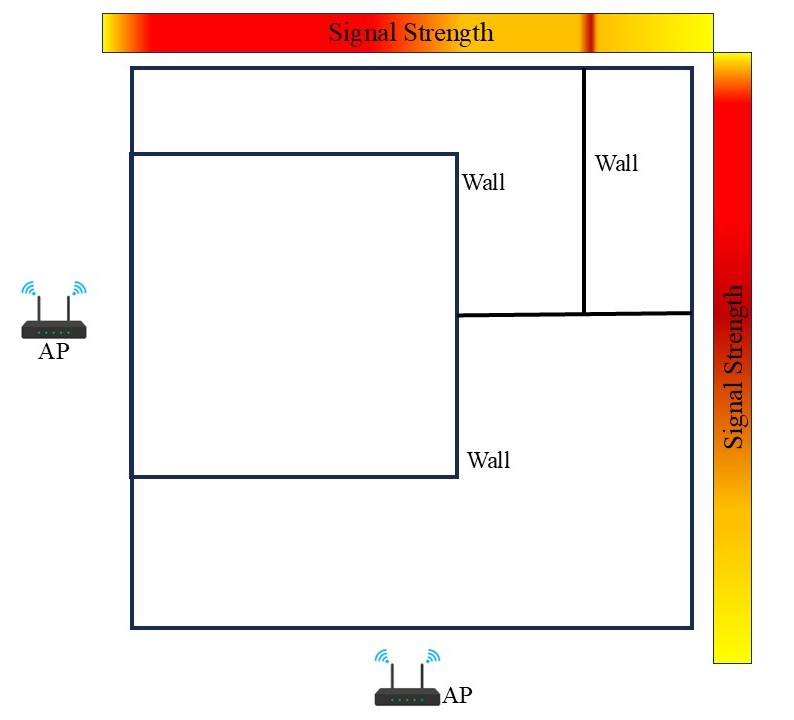} 
    \caption{RSSI Patterns for a Building}
    \label{RSSI patterns for a building}
\end{figure}

Passive reconnaissance of restricted places is possible since Wi-Fi is so widely used.
By positioning a receiver on one side of the building and access points on the other, this method can be used, for instance, to map a building in two dimensions. It is possible to estimate the building's walls and rooms using the signal levels that are produced. When mapping a building with several rooms, the RSSI values would appear as seen in Figure \ref{RSSI patterns for a building}. Outside the building, the APs were positioned outside the opposite wall, and the receivers were moved parallel to two walls. Furthermore, utilizing common smart home appliances like smart speakers or televisions, this may be utilized to install motion-activated devices, such as burglar alarms. 
 \section{Conclusion}
Wi-Fi could be leveraged to detect movement in confined areas by utilizing signal changes brought on by obstacles or motion as the system can function through radio transmission close to an AP. By utilizing the current wireless infrastructure, this creative method does not require extra hardware.
It is evident from our research that, human impediments and mobility have a substantial impact on RSSI values in LOS settings. The signal strength can drop by 10 to 15 dBm depending on the situation. The system demonstrated additional signal deterioration when impediments were introduced through walls, and it was able to detect brief disruptions with reliability, even while moving quickly. The distance between the obstacle and the AP or receiver affected the variations in RSSI, with larger decreases seen close to the AP. The system's adaptability was further demonstrated by the similar outcomes obtained by smart home appliances such as the Amazon Echo Dot. The findings of our study demonstrate how the technique can be applied to obstacle and motion detection.




\end{document}